\pdfoutput=1
\documentclass[10pt]{article}
\usepackage{amsmath,mathtools}
\DeclareMathOperator{\Tr}{Tr}
\usepackage{graphicx}
\usepackage{wrapfig}
\usepackage{enumitem, color, amssymb}
\usepackage{pgfplots}
\pgfplotsset{compat=1.12}
\usetikzlibrary{shapes,arrows,decorations.markings,positioning}
\usepackage{xr}
\usepackage{authblk}
\usepackage{hyperref}

\title{Three Dimensional Polarimetric Neutron Tomography of Magnetic Fields}

\author[1,*]{Morten Sales}
\author[2,3]{Markus Strobl}
\author[4]{Takenao Shinohara}
\author[5]{Anton Tremsin}
\author[6]{Luise Theil Kuhn}
\author[7]{William R.B. Lionheart}
\author[7]{Naeem M. Desai}
\author[8]{Anders Bjorholm Dahl}
\author[1,**]{S\o ren Schmidt}
\affil[1]{Department of Physics, Technical University of Denmark, DK-2800 Kgs. Lyngby, Denmark}
\affil[2]{Laboratory for Neutron Scattering and Imaging, Paul Scherrer Institute, 5232 Villigen, Switzerland}
\affil[3]{Niels Bohr Institute, University of Copenhagen, Copenhagen, DK-2100, Denmark}
\affil[4]{J-PARC Center, Japan Atomic Energy Agency, Tokai 319-1195, Japan}
\affil[5]{Space Sciences Laboratory, University of California at Berkeley, Berkeley, CA 94720, USA}
\affil[6]{Department of Energy Conversion and Storage, Technical University of Denmark, DK-4000 Roskilde, Denmark}
\affil[7]{School of Mathematics, The University of Manchester, Manchester, M13 9PL, United Kingdom}
\affil[8]{Department of Applied Mathematics and Computer Science, Technical University of Denmark, DK-2800 Kgs. Lyngby, Denmark}

\affil[*]{msales@fysik.dtu.dk}
\affil[**]{ssch@fysik.dtu.dk}

\begin{document}
\flushbottom
\maketitle
\begin{abstract}
\noindent Please find the now-published article\cite{Sales2018}: \\
Sales, M. et al. (2018). Three Dimensional Polarimetric Neutron Tomography of Magnetic Fields. {\it Scientific Reports}, {\bf 8(1)}, 2214. \\
at \url{https://doi.org/10.1038/s41598-018-20461-7}.\\

Through the use of Time-of-Flight Three Dimensional Polarimetric Neutron Tomography (ToF~3DPNT) we have for the first time successfully demonstrated a technique capable of measuring and reconstructing three dimensional magnetic field  strengths and directions unobtrusively and non-destructively with the potential to probe the interior of bulk samples which is not amenable otherwise.

Using a pioneering polarimetric set-up for ToF neutron instrumentation in combination with a newly developed tailored reconstruction algorithm, the magnetic field generated by a current carrying solenoid has been measured and reconstructed, thereby providing the proof-of-principle of a technique able to reveal hitherto unobtainable information on the magnetic fields in the bulk of materials and devices, due to a high degree of penetration into many materials, including metals, and the sensitivity of neutron polarisation to  magnetic fields.
The technique puts the potential of the ToF time structure of pulsed neutron sources to full use in order to optimise the recorded information quality and reduce measurement time.

\end{abstract}

\thispagestyle{empty}

\section*{Introduction}

The spin of a neutron passing through a magnetic field will undergo an amount of precession proportional to the strength of the magnetic field and the time spent by the neutron in the magnetic field. The time is proportional to the neutron wavelength, $\lambda$, and the path length through the magnetic field, $L$. The precession angle is given by\cite{Williams1988}:
\begin{align}
\phi=c \lambda B L,
\end{align}
where $c=4.632\times 10^{14}$~T$^{-1}$m$^{-2}$ is the Larmor constant, and $B$ is the magnetic field strength. Using this we can map the strength of a magnetic field along a neutron flight path into a neutron spin precession angle, and repeating this for multiple tomographic projections  we can reconstruct the magnetic field probed by the neutrons\cite{Hochhold1996}. 

In order to evaluate the potential of the technique we have chosen to measure the magnetic field generated by a current carrying solenoid, the magnetic field of which can be calculated for comparison, thereby providing the possibility for producing a solid proof-of-principle experiment investigating the capabilities of three dimensional magnetic field polarimetric neutron tomography (3DPNT). Previous experiments have successfully used monochromatic polarised neutrons beams for 2D imaging of magnetic fields\cite{Dawson2009}, as well as 
2D time-resolved imaging of periodically changing magnetic fields with a microsecond resolution\cite{Tremsin2015a}.
3D reconstructions with an assumption of the sample magnetic field direction to be along a direction perpendicular to the neutron polarisation has been demonstrated as well\cite{Kardjilov2008}. In contrast, our technique has been developed in order to measure and reconstruct 3D magnetic fields of arbitrary direction and distribution. This provides a method able to investigate samples without any {\it a priori} knowledge of the magnetic field orientation needed. Furthermore, it is able to use the full potential of a polychromatic pulsed neutron beam\cite{Strobl2009a}.

\section*{Experimental Set-up}
\begin{figure}[!ht]
\centering
\includegraphics[width=1\textwidth]{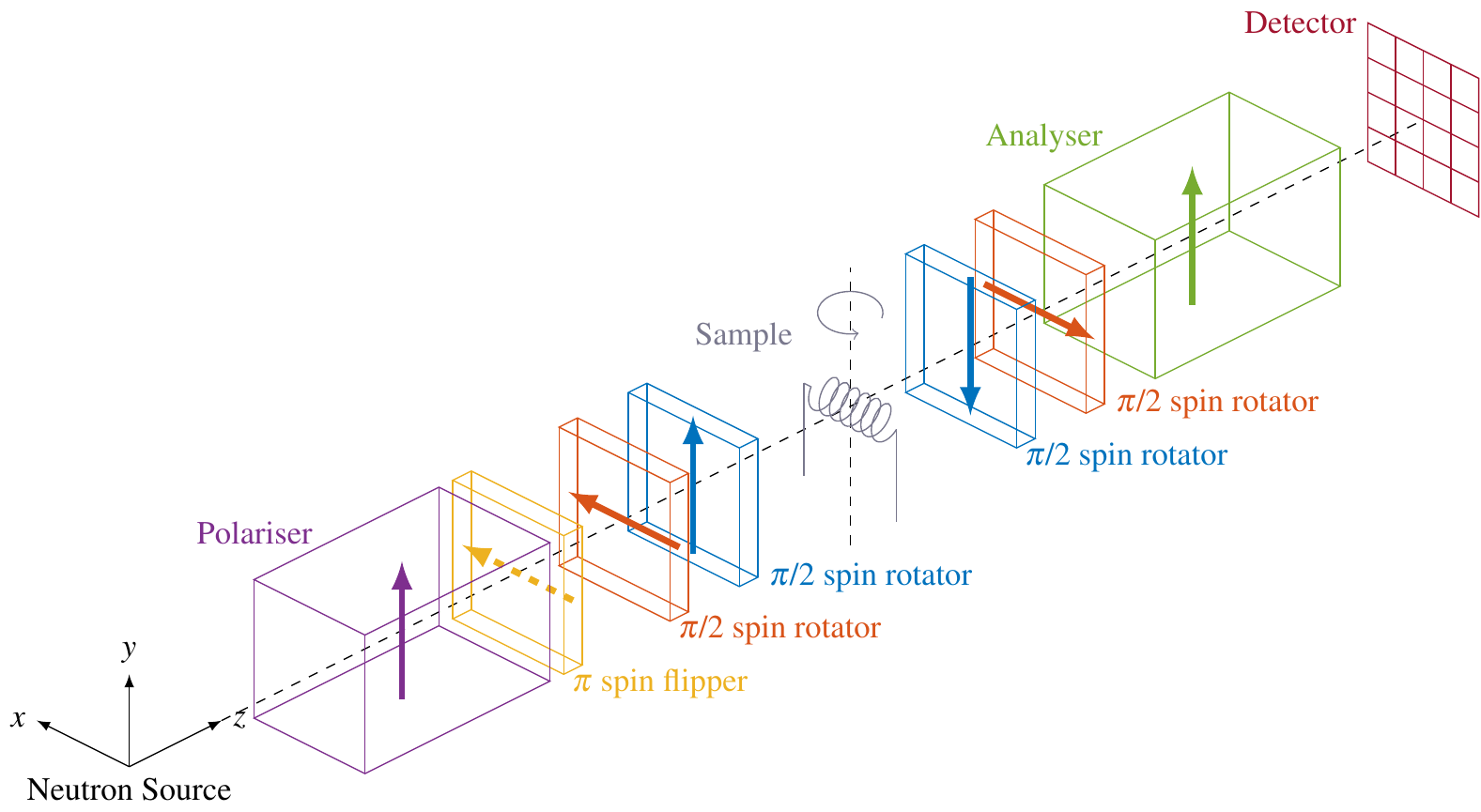}
\caption{Instrumental set-up. The neutrons are polarised in the $y$-direction by the polariser. The arrows of the  subsequent spin manipulators indicate the direction of the magnetic field around which the neutron spin is turned in the device. The $\pi$ spin flipper downstream from the polariser can be activated to select neutron polarisation directions of $-y$, and the two $\pi/2$ spin rotators further downstream can be used to rotate the neutron spins to the $\pm z$- or $\pm x$-directions. The sample is rotated around a vertical axis for different tomographic projections, and the two following $\pi/2$ spin rotators choose the direction of analysis before the analyser, which transmits neutrons with spins along $y$. Finally, the signal is recorded by a position and time sensitive detector.}
\label{fig:setup}
\end{figure}

To control the neutron spin direction before and after sample interaction, we use a polarimetric set-up where the initial neutron spin direction can be set to be parallel (or antiparallel) to either the $x$, $y$, or $z$ direction (see Fig.~\ref{fig:setup}). The neutron spin component along one of the same three axes can be analysed after the neutron has passed through the magnetic field of the sample\cite{Strobl2009b}.

Our experiments were performed at RADEN, BL22, at J-PARC MLF, Japan\cite{Shinohara2016}, with an instrumental set-up as described by Fig.~\ref{fig:setup}. Four $\pi /2$ spin rotators and a $\pi$ spin flipper are used to select directions of spin polarisation and analysis. The polariser and analyser are polarising supermirrors\cite{Shinohara2017} and a microchannel plate timepix detector\cite{Tremsin2015} with a $512\times 512$ array of $55\times 55$~$\mu$m$^2$ pixels and a temporal resolution of less than $1~\mu$s was used for neutron detection. We measured with a pulsed polychromatic neutron beam using the time-of-flight (ToF) information to determine the neutron wavelength, with the current in the spin rotators synchronised with the neutron pulse in order to achieve the proper neutron spin rotation for all neutron wavelengths. The time to wavelength conversion was done using a measurement of the Bragg edges of a standard iron sample, which provided the flight path and time delay values required for the conversion of ToF values into neutron wavelength. The images for all the wavelengths were acquired simultaneously and no scanning through energies was required in our set-up utilising the pulsed structure  of the neutron beam and high count rate capabilities of our ToF imaging detector. 
We rebinned our data to have a spatial binning of $10\times10$~pixels -- providing a spatial resolution of $\sim1$~mm -- and a temporal binning of $0.4992$~ms, corresponding to $\delta\lambda/\lambda=3.3\%$ at $\lambda=3.2$~{\AA}.

The sample used was an aluminium solenoid of length $L_s=1.55$~cm, radius $R=0.55$~cm, wire thickness of $w_t=0.1$~cm, with $N=13.5$~windings, and carrying a current of $I=0.6$~A. 

Neutron intensity data, $I_{\epsilon i, j}$ , for 60 projection angles between $0^{\circ}$ and $360^{\circ}$ was recorded with 18 different combinations of directions of spin polarisation, $\pm i$, and analysis, $j$, for each projection,  with $ i\in\{ z,y,z\}, \; j\in\{ x,y,z\}, \; \epsilon \in \{ -1,1\}$. The acquisition time for each of the $60\times18$ measurements was $\approx370$~s.








\section*{Reconstruction}
In order to reconstruct the measured magnetic field from the recorded polarimetric neutron tomographic data set, we have developed the reconstruction procedure presented in this section. As described in \cite{Leeb1998,Jericha2007} tomographic reconstruction of a magnetic field is not as straight forward as standard attenuation tomography since the polarisation of a neutron beam passing through a region of various magnetic field directions and strengths cannot be calculated using a simple line integral because of the non-commuting properties of the neutron spin orientation along the path\cite{Dawson2009}. Which is the reason we measured projections between 0$^{\circ}$ and 360$^{\circ}$, since neutron paths of opposite direction yield different outcomes. This can be exemplified by imaging a neutron with its spin direction along the $y$ direction passing through a magnetic field region with field direction along $x$, and of such extent and strength that it will rotate the neutron spin $90^{\circ}$ to be along the $z$ direction. A second magnetic field region of same size and strength and with the magnetic field direction along $y$ will further rotate the neutron spin to its final orientation along the $x$ direction. If the order of the two magnetic field regions had been switched, the final neutron spin direction would have been along the $z$ direction (the intermediate neutron spin direction being $y$). 
The polarisation was calculated from measured intensities as: 
\begin{align}
P_{i,j}=\frac{I_{(i,j)}-I_{(-i,j)}}{I_{(i,j)}+I_{(-i,j)}}\quad i\in\{ x,y,z\}, \; j\in\{ x,y,z\}.
\end{align} 
The effect of the magnetic field is cumulative along the ray and is governed by an ordinary differential equation\cite{Lionheart2015}. This means that the forward problem mapping the magnetic field, $\mathbf{B}$, to the measurements is not linear and so cannot be inverted with a simple inverse Radon transform. In this paper we apply a transformation of the data and then linearise the problem about $\mathbf{B}=0$. The measured intensities are of course scalars but the effect of the magnetic field is a rotation matrix  ${\bf P} \in SO(3)$ given by\cite{Strobl2009b}:

\begin{align}
{\bf P}=
\begin{bmatrix}
P_{x,x} & P_{x,y} & P_{x,z}\\
P_{y,x} & P_{y,y} & P_{y,z}\\
P_{z,x} & P_{z,y} & P_{z,z}
\end{bmatrix}.
\end{align}
An open beam measurement of $I_{y,y}$ and $I_{-y,y}$ was used to measure the flipping ratio (FR) and correct for the non-perfect polarisation and spin manipulation in the instrumental set-up. The FR was measured to be 22, corresponding to a polarisation of 91\%, averaged over the detector at $\lambda=3.2$~{\AA}.

In order to analyse the recorded data ${\bf P}$ was first transformed to a sample reference system.
For each projection, the sample was rotated around the vertical axis, $y$, and for a given projection angle, $\theta$, ${\bf R_y}(-\theta)$ is the rotation matrix:
\begin{align}
{\bf R_y}(-\theta)=
\begin{bmatrix}
\cos{(-\theta)} & 0 & \sin{(-\theta)}\\
0 & 1 & 0\\
-\sin{(-\theta)} & 0 & \cos{(-\theta)}
\end{bmatrix},
\end{align}
with which we can calculate the polarisation matrix, ${\bf P'}(\theta)$, in the sample reference system, $(x',y,z')$, using:
\begin{align}
{\bf P'}(\theta)={\bf R_y}(-\theta) \; {\bf P}(\theta) \; {\bf R_y}^\top(-\theta).
\end{align}
${\bf P'}$ is a rotation matrix that describes the spin rotation of a neutron caused by the magnetic field it travels through. 

The matrix exponential map takes skew symmetric matrices (the Lie algebra $\mathfrak{so}(3)$ ) to rotation matrices (the Lie group $SO(3)$). 
Geometrically for rotation matrix ${\bf P'}$  about an axis ${\bf \hat{k}}=(k_{x'}, k_{y}, k_{z'})$ through an angle $\phi$, the skew symmetric matrix is simply $\mathbf{K}\phi$ where
\begin{align}
\mathbf{K}=
 \begin{bmatrix} 0 & -k_{z'} & k_{y} \\ k_{z'} & 0 & -k_{x'} \\ -k_{y} & k_{x'} & 0 \end{bmatrix}  
\end{align}
and
\begin{align}
{\bf P'}=\exp(\phi \mathbf{K})  = \exp I + \sin \phi \mathbf{K} + (1- \cos \phi )\mathbf{K}^2
\end{align}
which is a matrix expression of the classical Rodrigues formula\cite{Rodrigues1840}. The inverse of this, the matrix logarithm is easily obtained. 
\begin{align}
\phi=\arccos{((\Tr({\bf P'})-1)/2)}=\arccos{((\Tr({\bf P})-1)/2)}
\end{align}
(with $\Tr$ denoting the matrix trace) and
\begin{align}
\mathbf{K}= ({\bf P'}-{\bf P'}^\top)/(2 \sin \phi).
\end{align}
Working in the logarithmic chart has the advantage that  $\mathfrak{so}(3)$ forms a vector space and the linearized forward problem takes the simple form in each plane of constant $y$  

\begin{align}
\mathcal{R} B_i = -\phi k_i/(c\lambda L) \quad i\in\{ x',y,z'\}
\end{align}
where $L$ is the (rebinned) pixel size and $\mathcal{R} $ is the two-dimensional Radon transform in the plane, which can be inverted by the standard filtered back projection methods\cite{Natterer2001} to obtain a reconstruction of the measured magnetic field described by $\mathbf{B}=(B_{x},B_{y},B_{z})$.

 A further limitation of the reconstruction algorithm is that it breaks down with phase wrapping, when the neutron spin precession angle gets larger than $180^{\circ}$. Since we normalise by wavelength we can average over all wavelengths where $\phi \leq 180^{\circ}$.

\begin{figure}
\centering
\includegraphics[width=1\textwidth]{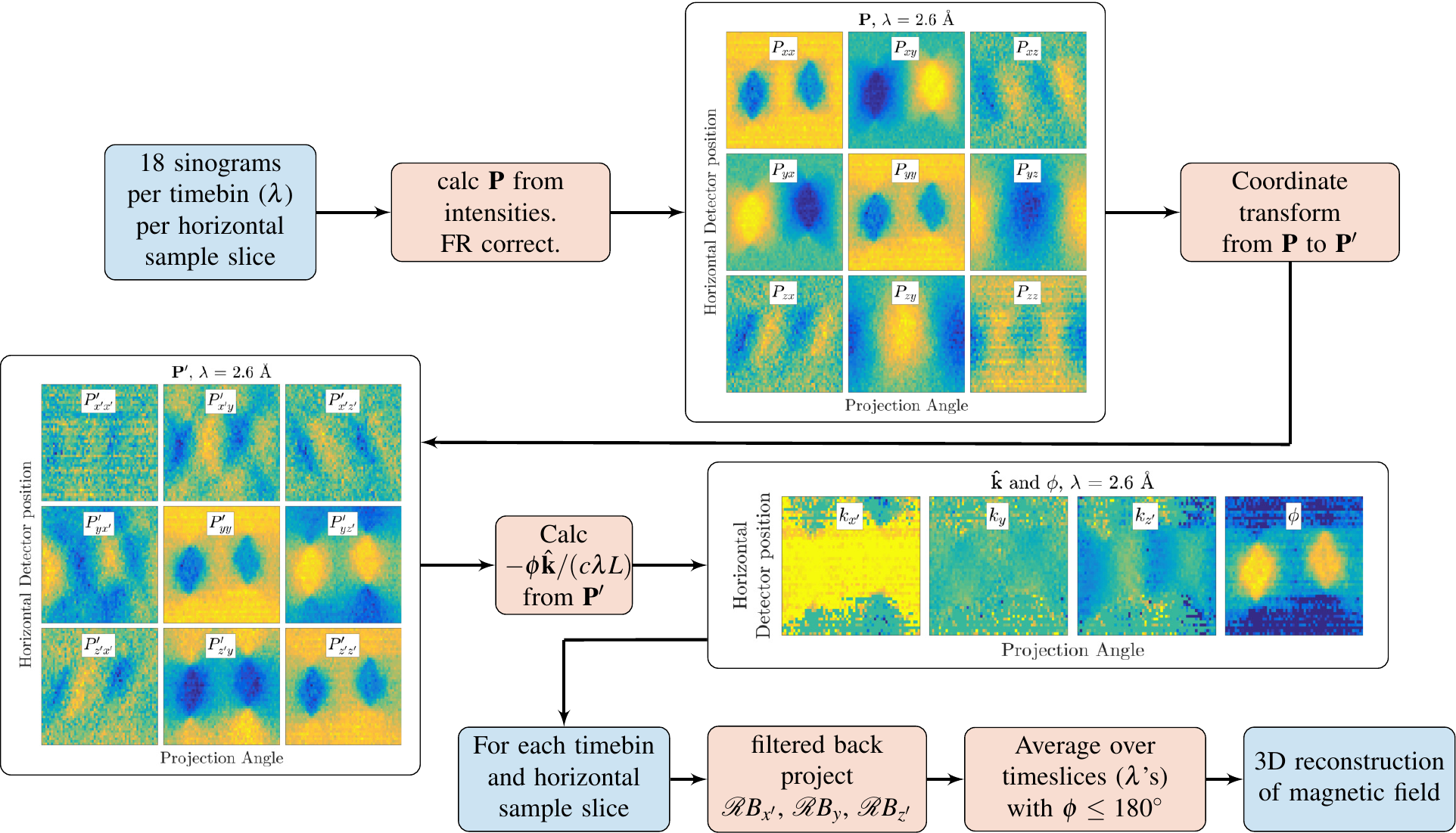}
    \caption{Procedure for reconstructing a 3D magnetic field measured with polarimetric neutron tomography. The measured intensities are reduced to 3 scalars that can be filtered back projected in order to obtain the reconstructed three dimensional magnetic field.}
    \label{fig:proc}
\end{figure}

The reconstruction procedure has been summed up in Fig.~\ref{fig:proc}, where examples of measured sinograms are shown.

\section*{Results}
\begin{figure}
\centering
\includegraphics[width=1\textwidth]{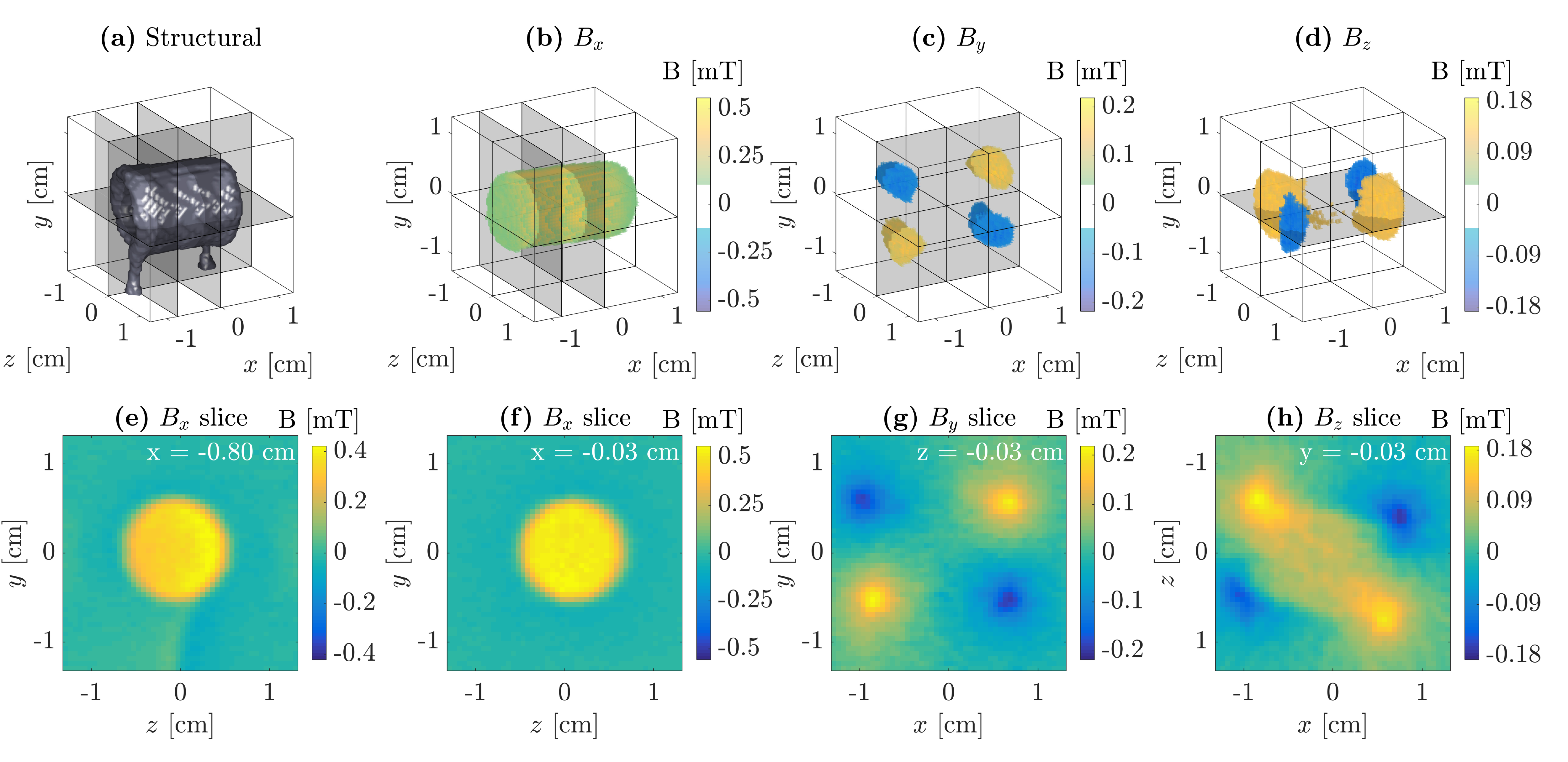}
\caption{\label{fig:recon1}Reconstruction results. {\bf (a)} shows the structural reconstruction from the attenuation signal. {\bf (b)}-{\bf (d)} respectively shows the $B_x$, $B_y$, and $B_z$ components of the reconstructed magnetic field for field strengths above the threshold indicated by the colorbars. {\bf (e)}-{\bf (h)} shows different slices through the reconstructed 3D magnetic field volume, with the slice location indicated by the gray planes in the above figures. Note that, as the solenoid was not oriented with its axis exactly along the x-axis, but rather at an $\sim 9^{\circ}$ angle in the $xz$-plane, a central signal can be seen along $z$ for the $B_z$ field in {\bf (h)}.
(Supplementary animation online at \cite{Sales2018}).}
\end{figure}

Figure~\ref{fig:recon1}~{\bf(b)}-{\bf(d)} show the $x$, $y$, and $z$ components of the reconstructed 3D magnetic field from the measured solenoid. It can be seen that the strongest field region along the solenoid axis is easily reconstructed as well as the weaker magnetic field areas where it "wraps around" the ends of the solenoid. Figure~\ref{fig:recon1}~{\bf(e)}-{\bf(h)} show selected slices (highlighted in {\bf(a)}-{\bf(d)}), where even the field from the current in the wires going to and from the solenoid is reconstructed as seen in {\bf (e)}.   

\begin{figure}
\centering
\includegraphics[width=1\textwidth]{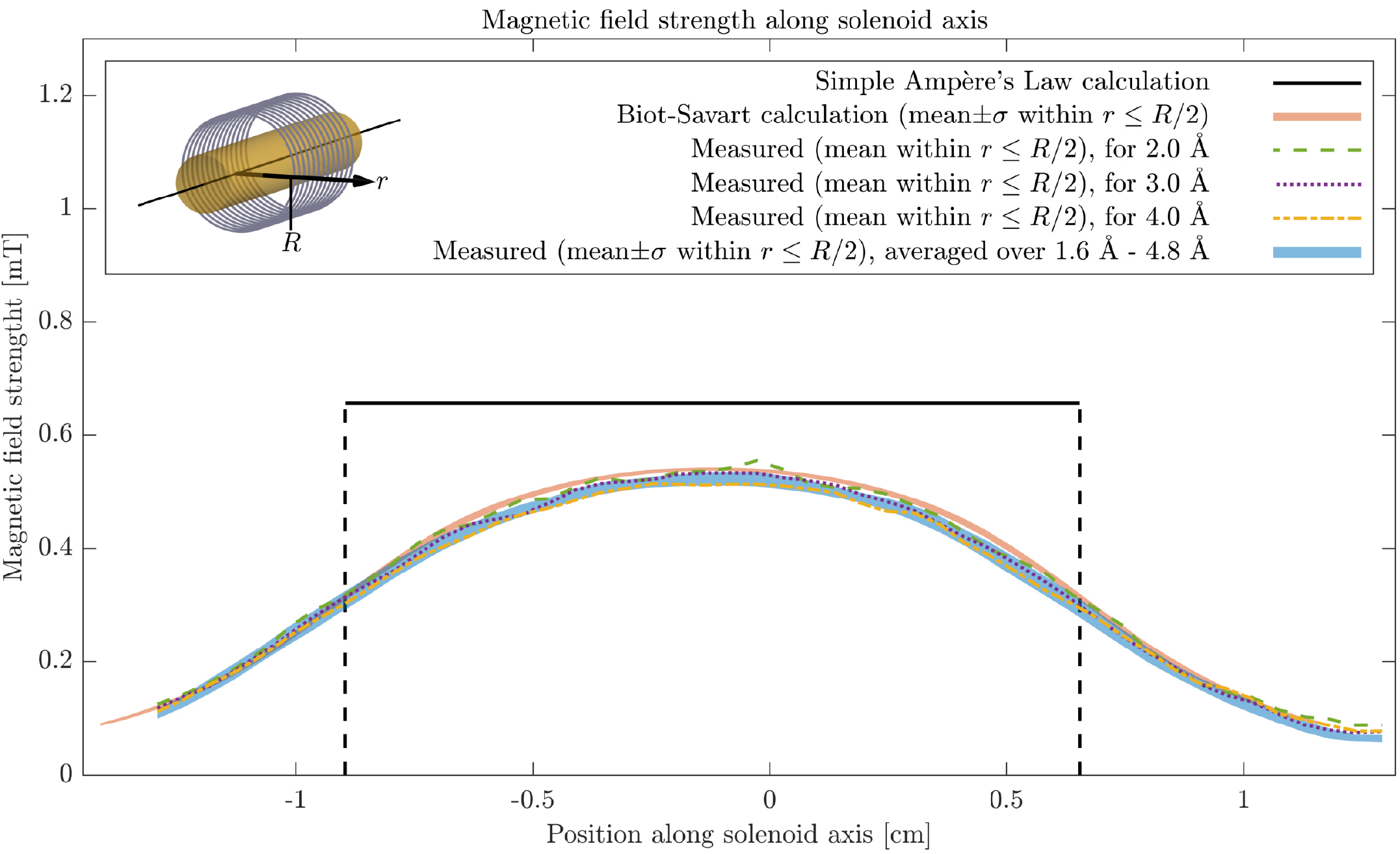}
\caption{\label{fig:reconaxes}Comparison between simple calculated magnetic field strength inside solenoid, a more precise calculation using the Biot-Savart law, and the reconstructed magnetic field strength using polarimetric neutron tomography. The field strength shown, as a function of position along the solenoid axis, is for the central cylindric area around the solenoid axis. The mean of the field as well as the standard deviation within this area is shown for the Biot-Savart calculation as well as the reconstructed field averaged over the measured wavelength range. Curves for the reconstructed field using three single wavelengths are included as well.}
\end{figure}

In order to compare our reconstruction to the expected resulting field from a current carrying solenoid, a calculation of the 3D magnetic field was done by dividing a description of the solenoid into 0.1~mm long straight wire segments and calculate the field contribution from each segment in a point cloud surrounding the solenoid using the Bio-Savart law:
\begin{align}\label{BS}
\mathbf {B} (\boldsymbol{\rho} )={\frac {\mu _{0}}{4\pi }}\;\sum _{\mathclap{\substack{\text{Wire}\\ \text{Segments}}}}\;{\frac {Id\mathbf {l} \times \boldsymbol{\rho'}}{|\boldsymbol{\rho'}|^{3}}},
\end{align}
where $\boldsymbol{\rho}$ is the point where the field is calculated, $\boldsymbol{\rho}'$ is the vector from the wire segment, $d{\bf l}$, to $\boldsymbol{\rho}$, and $\mu_{0}=4\pi\times 10^{-7}$~NA$^{-2}$ is the magnetic constant.
For further comparison a simple calculation using  Amp\`{e}re's Law ($B=\mu_0 I N /L$ ) was done as well. The results are shown in Fig.~\ref{fig:reconaxes}, where the magnetic field strength within the central part of the solenoid is shown. As expected the simple calculation overestimates the magnetic field strength. The same is true to a much smaller extend for the Bio-Savart calculation, though the small discrepancy between this and the reconstructed magnetic field from measurements can be attributed to imperfections in the measured solenoid. Also small imperfections in the instrumental set-up are not taken into account, as they are negligible compared to the polariser, analyser, and $\pi$ spin flipper efficiencies\cite{Shinohara2017}.
In the supplementary material (at \cite{Sales2018}), Fig.~A.2, a further comparison between measurements and calculation is shown as well as curves based on ray-tracing simulations.

\section*{Discussion}
We have with our successful measurements demonstrated the capabilities of a powerful technique for measuring three dimensional magnetic fields using ToF~3DPNT. Using the neutron spin precession in a magnetic field as a probe in combination with complicated reconstruction algorithms extracting information from the recorded data output, our proof-of-principle results are in a good agreement with calculations and serves as an initial demonstration of a novel technique that can extract hitherto unattainable information non-destructively from bulk samples.

In our FR correction we only corrected for the non-perfect polarisation characteristics of the polariser, analyser and $\pi$ spin flipper. It should be noted that in order to take into account the comparatively much smaller\cite{Shinohara2017} depolarisation in the spin rotators, further open beam measurements could have been performed at the expense of longer measurement time, or instead of using open beam measurements a polar decomposition by scaled Newton iteration could have been used to correct for small depolarisation effects thereby reorthonormalising ${\bf P'}$.

The current limitation of our technique is that the reconstruction relies on the assumption that we have linearised around $\mathbf{B}=0$ and that we have neutron spin precession angles below the phase wrapping limit ($\phi\leq180^{\circ}$).
If the phase wrapping limit lies within the measured wavelength band, it can be easily identified by following the progression of $\phi$ as the wavelengths increase, however, if the probed magnetic field is of such a strength that there is phase wrapping for even the fastest neutrons, the reconstruction algorithm would have to be expanded to possibly utilise the information contained in the period of $\phi$ as a function of wavelength\cite{Strobl2009a,Shinohara2011}. Furthermore, the wavelength band used can be adjusted to stay within the limits of the assumption of linearisation around $\mathbf{B}=0$.  To fully get around the assumption, an iterative reconstruction technique\cite{Leeb2013,Lionheart2015} with a forward model to approximate the measured field can be considered, as well as using vector field tomographic reconstruction\cite{Braun1991} on $\phi{\bf \hat{k}}$ directly (instead of using the assumption of linearisation around $\mathbf{B}=0$ to break it down to three scalars).

\section*{Outlook}
The unique information only obtainable with our novel method can be of use in a broad range of fields such as electrical engineering, superconductivity, energy materials, thermoelectrics, etc.

Combining the three dimensional magnetic field information with techniques providing structural information, such as conventional attenuation contrast imaging and more advanced methods\cite{Strobl2009c} like ToF three dimensional neutron diffraction (ToF~3DND)\cite{Cereser2017}, which can be performed using the same recorded data, 3DPNT provides a straight forward method for investigating the interplay between structural and magnetic sample composition.

\section*{Data Availability}
Data can be obtained from the authors by contacting Morten Sales (msales@fysik.dtu.dk).

%


\section*{Acknowledgements}
This work was supported by the European Union INTERREG \"Oresund-Kattegat-Skagerrak fund as well as DANSCATT. W.R.B.L acknowledges the Royal Society Wolfson Research Merit Award. The neutron experiments at the Materials and Life Science Experimental Facility at J-PARC, BL22, were performed under project number 2016I0022.

\section*{Additional information}
\subsection*{Author contributions statement}
S.S., M.St., T.S. and M.S. conceived the experiment. M.S, T.S., M.St., A.T., L.T.K., A.B.D, and S.S. conducted the experiment. M.S. and S.S. devised the reconstruction algorithm. M.S. and S.S. analysed the results. All authors reviewed the manuscript.

\subsection*{Competing Interests}
The authors declare no competing financial interests.

\end{document}